# Probing relaxation times in graphene quantum dots

Christian Volk[1,2,3]*, Christoph Neumann[1,2,3]*, Sebastian Kazarski[1,3], Stefan Fringes[1,3], Stephan Engels[1,2,3], Federica Haupt[3,4], André Müller[1,2,3], and Christoph Stampfer[1,2,3]

*1. II. Institute of Physics B, RWTH Aachen, 52074 Aachen, Germany*
*2. Peter Grünberg Institut (PGI-9), Forschungszentrum Jülich, 52425 Jülich, Germany*
*3. JARA – Fundamentals of Future Information Technologies*
*4. Institute for Theory of Statistical Physics, RWTH Aachen, 52074 Aachen, Germany*

*\*The first two authors contributed equally to this work.*

**Graphene quantum dots are attractive candidates for solid-state quantum bits. In fact, the predicted weak spin-orbit and hyperfine interaction promise spin qubits with long coherence times. Graphene quantum dot devices have been extensively investigated with respect to their excitation spectrum, spin-filling sequence, and electron-hole crossover. However their relaxation dynamics remain largely unexplored. This is mainly due to challenges in device fabrication, in particular regarding the control of carrier confinement and the tunability of the tunnelling barriers, both crucial to experimentally investigate decoherence times. Here, we report on pulsed-gate transient spectroscopy and relaxation time measurements of excited states in graphene quantum dots. This is achieved by an advanced device design, allowing to tune the tunnelling barriers individually down to the low MHz regime and to monitor their asymmetry with integrated charge sensors. Measuring the transient currents through electronic excited states, we estimate lower limit of charge relaxation times on the order of 60–100 ns.**

Since the seminal work by Loss and DiVincenzo [1], quantum dots (QDs) are intensively tested as building blocks for solid-state quantum computation. Today, the most advanced implementations of QD qubits are realized in III/V heterostructures [2-4]. Nevertheless, the strong hyperfine interaction in these compounds poses fundamental limits to the spin coherence time, stimulating the search for alternative host materials, especially within the group IV elements [5]. In this context, graphene is a particularly promising candidate, thanks to its low nuclear spin densities and weak spin-orbit interaction [6-8]. However, the gapless electronic band structure and the Klein tunnelling phenomena [9] make it challenging to confine electrons electrostatically. This limitation can be circumvented by etching nanostructures in graphene, thus introducing a disorder-induced energy gap that allows confining individual carriers [10-17]. So far, excitation spectrum [18-21], spin-filling sequence [22], and electron-hole crossover [23] have been studied in graphene QDs. However, these devices lack the tunability of GaAs devices, in particular with respect to the transparency of the tunnelling barriers, hindering the possibility to address the relaxation dynamics of individual quantum states experimentally.

Here, we present QDs with highly tunable barriers formed by long and narrow constrictions. This design is motivated by our recent study on the aspect ratio of graphene nanoribbons [17], and it proved optimal to realize highly tunable tunnelling barriers, down to the low MHz regime. This enables us to perform pulsed-gate excited state spectroscopy and to give a first estimate of relaxation times in graphene QDs.

Our devices consist of an etched graphene island (diameter, $d = 110$ nm) connected to source and drain leads by long and narrow constrictions serving as tunnelling barriers (Figs. 1a,b). Two graphene nanoribbons, located symmetrically on both sides of the island, can be simultaneously used as gates (LG, RG) to tune the transparency of the barriers and as detectors sensitive to individual charging events in the dot [24,25]. The electrostatic potential of the QD is controlled by a central gate (CG), on which a bias-tee mixes AC and DC signals (Fig. 1b). This allows performing pulsed-gate experiments with the same gate used for DC control. The devices are tuned by a back gate in the low charge-carrier density regime, where transport is dominated by Coulomb blockade effects [18-20] (*cf.* Supplementary material).

The QD is first characterized via low-bias transport measurements (Fig. 1c) as a function of the voltage applied to the gates that control the transparency of the barriers [18]. The optimized design of our device allows us to tune individually the tunnelling rates of both barriers $\Gamma_L$, $\Gamma_R$ over a wide range. Suppressing these rates, we access the sequential tunnelling regime. Figure 1d shows two simultaneous measurements of the current through the dot (lower curve) and through the nanoribbon on its right side (upper curve) as a function of the voltage $V_{CG}$ applied to the central gate CG. The dot current exhibits equally spaced Coulomb resonances, with peak currents up to a few hundred fA. Note that a peak current of *e.g.* 200 fA corresponds to an overall dot-transmission rate of $\Gamma < 1.25$ MHz. In correspondence to each peak, the current in the nanoribbon shows characteristic sharp kinks, which indicate that the latter acts as a sensitive charge detector (CD) for the quantum dot [24,25].

The addition energy of the dot and its excited state spectrum are probed by finite-bias spectroscopy. Figures 2a,b show the current and the differential conductance through the QD as a function of the applied bias $V_{SD}$ and of the gate voltage $V_{CG}$. These Coulomb diamond measurements give an estimate of the addition energy of the QD, $E_{add} \approx 10.5$ meV. In a simple disc-capacitor model, this corresponds to a QD diameter of 120 nm, in good agreement with the geometric size of our device. Clear signatures of transport through well-defined excited states can be observed already in the current (see *e.g.* Fig. 2b) and become even more evident in the differential conductance Fig. 2c, from which we extract a level spacing of about $\Delta = 1.5 - 2.5$ meV, in agreement with the electronic single-particle level spacing given by $\Delta = \hbar v_F/(d\sqrt{N}) \approx 1.3 - 1.9$ meV, where $N$, the number of carriers on the dot, is assumed to be on the order of 10-20 [20].

Simultaneously to the current through the dot, we measure the one flowing through the charge detector $I_{CD}$. Its derivative with respect to the gate voltage $V_{CG}$ is shown in Fig. 2d. Regions of large $dI_{CD}/dV_{CG}$ signal the onset of those transitions that most affect the occupation of the QD. If they appear in concomitance of both edges of a certain Coulomb diamond (*cf.* diamond centred around $V_{CG} \approx 4.2$ V), this indicates a regime in which the coupling to the two leads is of comparable magnitude. *Vice versa*, regimes where one of the tunnel barriers is dominating are characterized by a strongly one-sided $dI_{CD}/dV_{CG}$ (*cf.* diamond centred around $V_{CG} \approx 4.45$ V). In this way, we can monitor the asymmetry of the tunnel barriers. Together with their good tunability, these permit to identify regimes where to perform pulsed-gate transient spectroscopy of excited states, *i.e.* regimes where *both* tunnel rates $\Gamma_L$, $\Gamma_R$ are much smaller than the inverse pulse rise-time $\tau_{rise}^{-1}$. The pulse-gating technique relies in fact crucially on the pulse rise-time $\tau_{rise}$ being the fastest time scale in the system. Conductance measurements in such a regime are shown *e.g.* in Fig. 2e. Note that the charge detector is such a sensitive electrometer, that QD excited states can be clearly resolved in the differential transconductance $dI_{CD}/dV_{SD}$ (Fig. 2f).

To perform pulsed-gate spectroscopy of QD excited states, we mix a square-shaped AC signal (see Fig. 3a) to the DC voltage $V_{CG}$ applied to the central gate and measure DC-transport through the dot for small (fixed) $V_{SD}$. If the frequency of the pulse is low (100 kHz in Figs. 3b,c), this simply results in the splitting of a Coulomb resonance in two peaks [26]. These stem from the QD ground state (GS) entering the bias-window at two different values of $V_{CG}$, one for the lower pulse level (A), and one for the upper (B). At larger frequencies (800 kHz in Fig. 3d), together with the splitting there is an additional broadening of the peaks. The situation changes dramatically at much higher frequencies (tens of MHz in Figs. 3e,g), where a number of additional peaks due to transient transport through QD excited states (ES) appear. Each of these resonances corresponds to situations in which the QD level is pushed well outside the bias-window in the first half of the pulse, and then brought in a position where transport can occur only via excited states, see *e.g.* Fig. 3f. A transient current can flow in this second half of the pulse until some process causes the occupation of the GS, causing its blockade. This current can be resolved in our DC-measurements only if the frequency of the pulse is higher than the characteristic rate $\gamma$ of the blocking processes (see also supplementary material). Note that the energies of the ES extracted via pulsed-gate spectroscopy are in excellent agreement with those determined from DC Coulomb diamond measurements, Figs. 3e,h.

A more accurate analysis of transient currents via ES can be obtained with a different pulse scheme (see Fig. 4a), where $T_A$ is varied while keeping $T_B$ and $V_{PP}$ fixed [27-29]. Such a measurement is shown in Fig. 4b for the same Coulomb resonance investigated in Figs. 3b-e. The two outermost peaks correspond to transport via the dot GS when this is in resonance with the bias window during $T_A$ (right peak) and $T_B$ (left peak). The inner peak results from transport via an excited state, according to the scheme sketched in Fig. 4c. For each of these peaks, we estimate the average number of electrons tunnelling through the device per cycle $\langle n \rangle = I (T_A+T_B)/e$. In Fig. 4d, we plot the number of electrons transmitted via the GS (blue) and via the ES (red) as a function of the pulse length $T_A$. As expected, while the first one increases linearly with $T_A$, the second tends to saturate, indicating a transient effect. Fitting this data-set with $n(T_A) = n_{sat} [1-exp(-\gamma T_A)]$, where $n_{sat}$ is the saturation value for long $T_A$ [28], we extract the characteristic rate $\gamma = 12.8$ MHz of the blocking processes. As both tunnelling and relaxation concur to the occupation of the GS, $\gamma$ is approximately given by $\gamma \sim \Gamma + 1/\tau$, where $\tau$ is the relaxation time of the excited state and $\Gamma$ the characteristic tunnelling rate ($\Gamma \sim 2.5$ MHz in this case). This in turn gives a lower bound $\tau > 78$ ns for the life-time of the excited dot states. By studying more electronic excited states, all with energies in the range of 1.7 to 2.5 meV, we estimate a lower bound for the relaxation time in the range of 60 - 100 ns. This timescale is likely related to the lifetime of charge excitations, which is limited by the electron-phonon interaction. The extracted charge relaxation times are a factor 5-10 longer than what has been reported in III/V quantum dots [27-30]. This might be a signature of reduced electron-phonon interaction in $sp^2$-bound carbon, where piezoelectric phonons are absent. However, an accurate estimate of the electron-phonon interaction in etched graphene quantum dots is difficult, since the destroyed graphene lattice leads to a phonon density of states that is strongly dependent on the exact shape of the dot itself.

In summary, we perform transient-current spectroscopy of excited states in graphene QDs, obtaining a lower bound of about $60 - 100$ ns for the relaxation time of electronic excitations. This experiment represents a fundamental step towards the investigation of spin life times and coherence times in graphene quantum dots.

## Methods

The sample is fabricated by mechanical exfoliation of natural bulk graphite. Graphene flakes are transferred to highly doped silicon substrates with a 295 nm silicon oxide top-layer. Single-layer flakes are selected via Raman spectroscopy (see also suppl. material), and then patterned by electron beam lithography (EBL) followed by $Ar/O_2$ reactive ion etching. A second EBL and lift-off step is performed to place source and drain electrodes and gate contacts (all 5/50 nm Cr/Au) on the device. Measurements are performed in a dilution refrigerator, at a base temperature around 20 mK. Home-built low-noise DC amplifiers are used to measure current with precision below 100 fA. For pulsed-gate experiments, we use the arbitrary waveform generators Tektronix AWG520 and Tektronix AWG7082C. The rise-time detected close to the sample is about 250-500 ps. The bias-tee Anritsu K251 mixes AC and DC signals.

1. Loss, D., DiVincenzo, D. P. Quantum computation with quantum dots. *Phys. Rev. A* **57**, 120 (1998).

2. Petta, J. R., *et al.* Coherent manipulation of coupled electron spins in semiconductor quantum dots. *Science* **309,** 2180 (2005).

3. Nowack, K. C. *et al.* Single-Shot Correlations and Two-Qubit Gate of Solid-State Spin. *Science* **333,** (6047) 1269 (2011).

4. Shulman M. D. *et al.* Demonstration of Entanglement of Electrostatically Coupled Singlet-Triplet Qubits. *Science* **336**, 202 (2012).

5. Hu, Y., Kuemmeth, F., Lieber, C.M., Marcus, C.M. Spin relaxation in Ge/Si Core-Shell Nanowire Qubits. *Nat. Nano.***7**, 47 (2011).

6. Trauzettel, B., Bulaev, D., Loss, D., Burkard, G. Spin qubits in graphene quantum dots. *Nature Physics* **3**, 192 (2007).

7. Min, H. *et al.* Intrinsic and Rashba spin-orbit interactions in graphene sheets. *Phys. Rev. B* **74**,165310 (2006).

8. Huertas-Hernando, D., Guinea, F., Brataas, A. Spin-orbit coupling in curved graphene, fullerenes, nanotubes, and nanotube caps. *Phys. Rev. B* **74**, 155426 (2006).

9. Katsnelson, M.I., Novoselov, K.S., Geim, A.K. Chiral tunnelling and the Klein paradox in graphene. *Nat. Phys.* **2**, 620 (2006).

10. Han, M. Y., Özyilmaz, B., Zhang, Y., Kim, P. Energy Band-Gap Engineering of Graphene Nanoribbons. *Phys. Rev. Lett.* **98**, 206805 (2007).

11. Stampfer, C., Güttinger, J., Hellmüller, S., Molitor, F., Ensslin, K., Ihn, T. Energy gaps in etched graphene nanoribbons. *Phys. Rev. Lett.* **102,** 056403 (2009).

12. Todd, K., Chou, H.-T., Amasha, S., Goldhaber-Gordon, D. Quantum Dot Behavior in Graphene Nanoconstrictions. *Nano Lett.* **9**, 416 (2009).

13. Liu, L., Oostinga, J.B., Morpurgo, A.F., Vandersypen, L.M.K. Electrostatic confinement of electrons in graphene nanoribbons. *Phys. Rev. B* **80**, 121407 (2009).

14. Ihn, T. *et al.* Graphene single-electron transistors. *Materials Today* **13**, 44 (2010).

15. Gallagher, P., Todd, K., Goldhaber-Gordon, D. Disorder-induced gap behavior in graphene nanoribbons. *Phys. Rev. B* **81**, 115409 (2010).

16. Han, M.Y., Brant, J.C., Kim, P. Electron Transport in Disordered Graphene Nanoribbons. *Phys. Rev. Lett.* **104**, 056801 (2010).

17. Terres, B., Dauber, J., Volk, C., Trellenkamp, S., Wichmann, U., Stampfer, C. Disorder induced Coulomb gaps in graphene constrictions with different aspect ratios. *Appl. Phys. Lett.* **98,** 032109 (2011).

18. Stampfer, C., Schurtenberger, E., Molitor, F., Güttinger, J., Ihn, T., Ensslin, K. Tunable graphene single electron transistor. *Nano Lett.* **8,** 2378 (2008).

19. Ponomarenko, L. A. *et al.* Chaotic Dirac Billiard in Graphene Quantum Dots. *Science* **320**, 356 (2008).

20. Schnez, S. *et al.* Observation of excited states in a graphene quantum dot. *Appl. Phys. Lett.* **94**, 012107 (2009).

21. Moser, J., Bachtold, A. Fabrication of large addition energy quantum dots in graphene. *Appl. Phys. Lett.* **95**, 173506 (2010).

22. Güttinger, J., Frey, T., Stampfer, C., Ihn, T., Ensslin, K. Spin states in graphene quantum dots. *Phys. Rev. Lett.* **105**, 116801 (2010).

23. Güttinger, J. *et al.* Electron-Hole Crossover in Graphene Quantum Dots. *Phys. Rev. Lett.* **103**, 046810 (2009).

24. Güttinger, J. *et al.* Charge detection in graphene quantum dots. *Appl. Phys. Lett.* **93,** 212102 (2008).

25. Wang, L.-J., *et al.* A graphene quantum dot with a single electron transistor as an integrated charge sensor. *Appl. Phys. Lett.* **97**, 262113 (2010).

26. Dröscher, S. *et al.* High-frequency gate manipulation of a bilayer graphene quantum dot. *Appl. Phys. Lett.* **101**, 043107 (2012).

27. Fujisawa, T., Austing, D.G., Tokura, Y., Hirayama, Y., Tarucha, S. Nonequilibrium Transport through a Vertical Quantum Dot in the Absence of Spin-Flip Energy Relaxation. *Phys. Rev. Lett.* **88**, 236802 (2002).

28. Fujisawa, T., Tokura, Y., Hirayama, Y. Transient current spectroscopy of a quantum dot in the Coulomb blockade regime. *Phys. Rev. B,* **63**, 081304 (2001).

29. Fujisawa, T., Austing, D.G., Tokura, Y., Hirayama, Y., Tarucha, S. Allowed and forbidden transitions in artificial hydrogen and helium atoms. *Nature* **419,** 278 (2002).

30. Jang, Y.D. *et al.* Carrier lifetimes in type-II InAs quantum dots capped with a GaAsSb strain reducing layer. *Appl. Phys. Lett.* **92**, 251905 (2008).

Acknowledgements - The authors wish to thank Stefan Trellenkamp for electron beam lithography and Uwe Wichmann for low-noise measurement electronics. We thank Hendrik Bluhm, Guido Burkard, Fabian Hassler, Thomas Ihn and Markus Morgenstern for helpful discussions. Support by the DFG (SPP-1459 and FOR-912) and the ERC are gratefully acknowledged.

Correspondence and requests for materials should be addressed to C.S. (stampfer@physik.rwth-aachen.de).

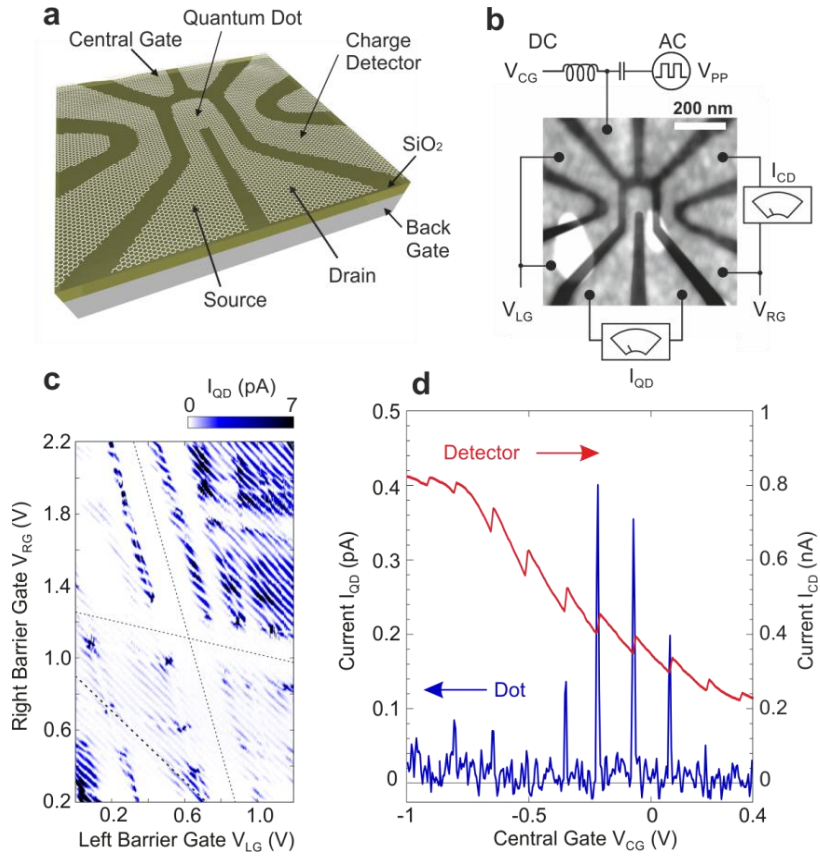

Figure 1: **Graphene quantum dot device and low bias transport measurements. a-b**, Schematic and scanning force micrograph of the measured device. The central island is connected to source (S) and drain (D) electrodes by two 40 nm wide and 80 nm long constrictions, whose transparency can be tuned by the voltage applied to the nearby nanoribbons ($V_{LG}$, $V_{RG}$). The right nanoribbon is also used as a charge detector (CD) for the dot. The central gate CG is connected to a bias-tee mixing AC and DC signals. **c,** Current through the dot as a function of the barrier gate-voltages $V_{LG}$, $V_{RG}$, recorded at a source-drain voltage $V_{SD}$ = -1.5 mV and $V_{CG}$ = 0 mV. Here and in the following, the electron temperature is $T_e$ < 100 mK, and the back gate voltage is $V_{BG}$ = 34.6 V, corresponding to a Fermi level position deep in the transport gap (*cf.* supplementary material). Different families of resonances can be identified in this measurement, which can be either attributed to the dot (features with relative lever arm of 0.9; dotted line) or to localized states either in the left or right constriction (features with a relative lever arm of 5 and of 0.25, respectively; dashed lines). This indicates the possibility of tuning the transparency of the two constrictions independently, and to control the current through the dot down to the sub-pA level by acting on $V_{LG}$, $V_{RG}$. **d,** Simultaneous measurement of the current flowing through the dot and through the right nanoribbon as a function of $V_{CG}$. The bias voltage applied to the dot and to the nanoribbon are $V_{SD}$ = -1.5 mV and $V_{CD}$ = 0.2 mV, respectively. Barrier-gate voltages are $V_{LG}$ = 0.4 V, $V_{RG}$ = 0 V.

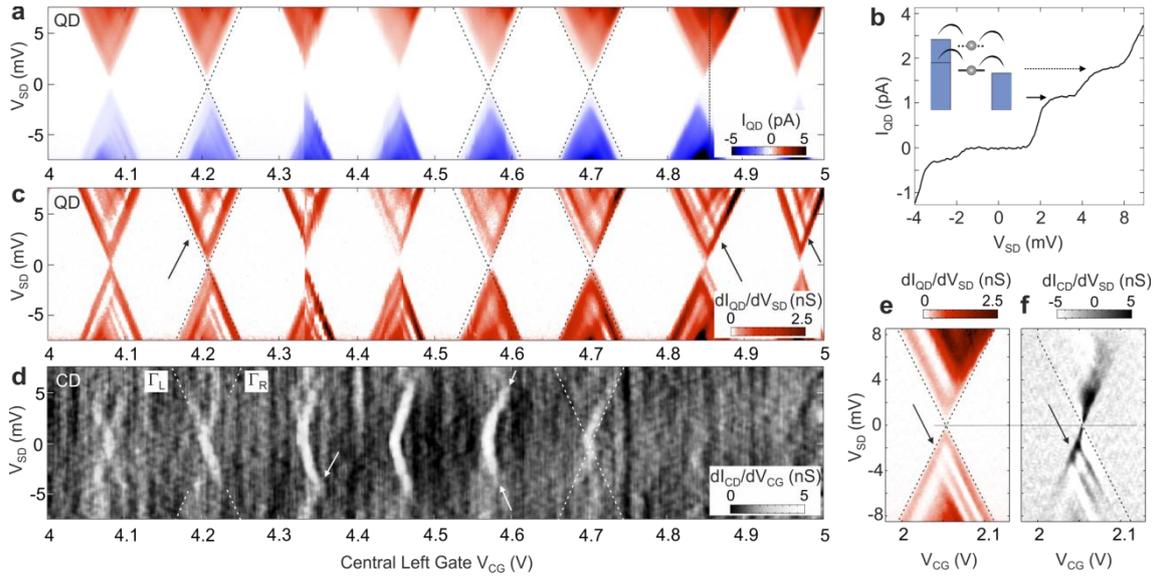

Figure 2: **Finite-bias excited states spectroscopy and charge sensing. a.** Current through the quantum dot as function of $V_{CG}$ and $V_{SD}$ in a regime of weak tunnel coupling to leads; the dashed lines are guides to the eye indicating the edges of the Coulomb diamonds. **b.** Line-cut along the line of constant $V_{CG}$ indicated in (a): the stepwise increase of the current is a signature of the discrete QD spectrum; the two well-defined plateaus correspond to the GS and the first ES respectively being included into the bias-window. **c.** Differential conductance of the QD, $dI_{QD}/dV_{SD}$. Resonances parallel to both edges of the Coulomb diamonds, indicating transport via excited states, can clearly be seen. **d.** Derivative of the charge-detector current $I_{CD}$ with respect to $V_{CG}$. Regions of high $dI_{CD}/dV_{CG}$ correspond to the onset of the transitions with the largest rate, thus providing information on the asymmetry of the tunnelling barriers. Note that this can be bias dependent (*e.g.* the diamond centred on $V_{CG} \approx 4.45$ V is more strongly coupled to the right lead for $V_{SD} > 0$ and to the left one for $V_{SD} < 0$) and it can be drastically influenced by the onset of transitions via excited states, as indicated by the appearance of kinks as those shown by the arrow. **e,f.** Simultaneous measurements of the differential conductance of the dot $dI_{QD}/dV_{SD}$ (e) and of the transconductance of the charge detector $dI_{CD}/dV_{SD}$ (f) in a regime of interest for pulsed-gate experiments.

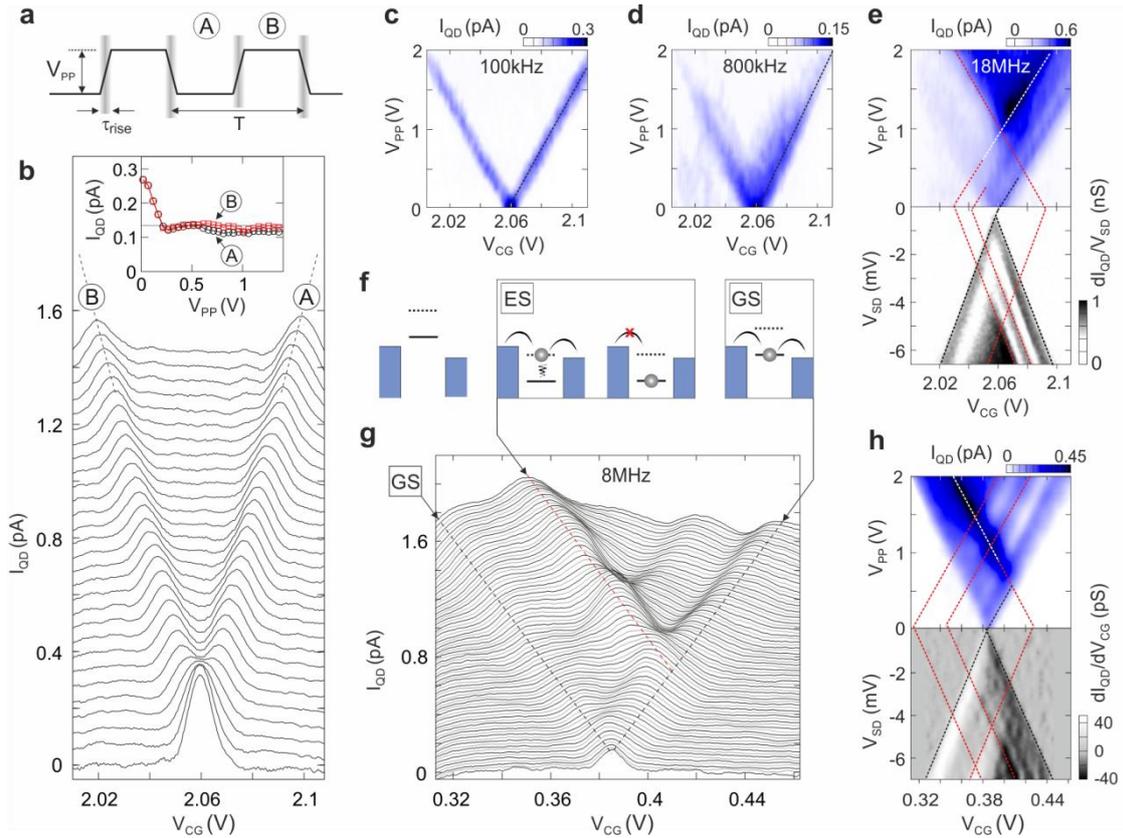

Figure 3: **Pulsed-gate spectroscopy of excited states. a.** Sketch of the pulse scheme employed in all measurements presented in this figure (duty cycle 50%). Low and high pulse level are labelled by A and B, respectively. **b.** Current through the dot under the influence of a 100 kHz pulse. Different lines correspond to $V_{PP}$ being varied from 0 to 1.4 V in steps of 50 mV (lines offset by 0.05 pA). Here and in the following $V_{SD}$ = -1.5 mV. Increasing the amplitude, the Coulomb peak splits in two resonances that shift linearly with pulse amplitude and whose height is approximately half of the original height (see inset). **c-d.** Same measurement as in (b), but presented as a colour-scale plot. In (d) the pulse frequency is 800 kHz. **e.** Upper panel: same measurement as above, but at a higher frequency of 18 MHz. Together with the ground state splitting, a number of additional resonances can be seen, corresponding to transient currents via excited states. The level splitting extracted from this measurement coincides with the one given by DC finite-bias measurements (lower panel, same data as in Fig. 2e). The dashed lines are guides to the eye. **f.** Schematic of transport via ground state (GS), excited state (ES) and, on the left, of a possible initialization stage. **g.** Measurement similar to the one shown in (b), but for a different Coulomb resonance. Here the pulse frequency is 8 MHz, $V_{SD}$ = -1 mV, and $V_{PP}$ is varied from 0 to 2 V in steps of 25 mV (lines offset by 0.02 pA). **h.** Same data set as in (g), and comparison with the corresponding DC measurement, showing excellent agreement between the excited states level spacing extracted from pulsed-gate and finite-bias spectroscopy.

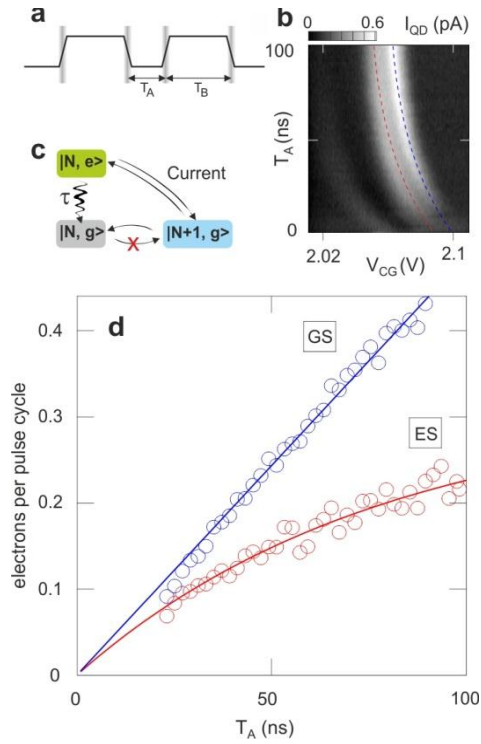

Figure 4: **Transient current through excited states. a.** Pulse scheme used to study transient current through excited states: the duration of $T_B$ is kept constant, while $T_A$ is varied. **b.** Current through the QD as a function of $V_{CG}$ and of the time $T_A$ for $V_{SD} = -1$ mV, $V_{PP} = 0.9$ V, $T_B = 30$ ns. All lines are bent due to a parasitic effect of the bias-tee adding a duty-cycle dependent DC-offset to $V_{CG}$. Dashed lines indicate the peaks due to transport via ground (blue) and excited (red) states further investigated in (d). **c.** Schematic of the transitions occurring during $T_A$ for the middle peak in (b). $|N,g\rangle$ and $|N,e\rangle$ indicate the ground and the excited state of the dot with N excess electrons, respectively. The transition $|N,g\rangle \rightarrow |N+1,g\rangle$ is not energetically allowed for these values of $V_{SD}$ and $V_{CG}$ during $T_A$, making $|N,g\rangle$ an absorbing state. In each cycle, the dot is initialized in the state with N+1 electrons during $T_B$ (not shown). **d.** Average number of electrons transmitted per cycle via GS (blue circles) and ES (red circles), as a function of the pulse length $T_A$. Data are extracted from the peaks shown in (b), solid lines represent linear and exponential fits to the experimental data ($1/\gamma = 78$ ns, $n_{sat} = 0.313$).

# Supplementary Information for:

# Probing relaxation times in graphene quantum dots

Christian Volk[1,2,3]*, Christoph Neumann[1,2,3]*, Sebastian Kazarski[1,3], Stefan Fringes[1,3], Stephan Engels[1,2,3], Federica Haupt[3,4], André Müller[1,2,3], and Christoph Stampfer[1,2,3]

1. II. Institute of Physics B, RWTH Aachen, 52074 Aachen, Germany
2. Peter Grünberg Institut (PGI-9), Forschungszentrum Jülich, 52425 Jülich, Germany
3. JARA – Fundamentals of Future Information Technologies
4. Institute for Theory of Statistical Physics, RWTH Aachen, 52074 Aachen, Germany

*The first two authors contributed equally to this work.

## 1. Sample characterisation

After exfoliation of the graphene flakes, Raman spectroscopy was employed to identify single layer flakes [s1]. The Raman spectrum of the measured graphene flake shows a 2D-peak with FWHM of 34 cm$^{-1}$, see Fig. S1a. The peak amplitude of the 2D-line is almost twice as high as the G-line proving that the flake is one monolayer thin.

Figure S1b shows the back-gate characteristics of our graphene quantum dot. The back-gate affects the overall Fermi level of the sample, thus allowing tuning the device both into the electron and the hole transport regimes, as well as in the transport gap [s2]. All measurements shown in the main text have been performed deep inside the transport gap, at the back gate voltage of $V_{BG}$ = 34.6 V (see arrow).

## 2. Pulsed-gate spectroscopy on different resonances

In Figures S2 and S3, we present additional examples of different Coulomb resonances under the influence of pulsed-gating. The additional data shown in Fig. S2 are supplementary to the ones presented in Fig. 3 in the main text, showing the evolution as a function of the pulse amplitude of the same resonances considered there, but at different pulse frequencies. The results are qualitatively independent of the frequency, indicating that this is faster than the characteristic blocking rate γ. Note that the pulse frequency of 8 MHz in Fig. S3d is lower than the estimate for γ obtained from the measurements presented in Fig. 4 in the main text. This discrepancy is due to incomplete initialization of the dot during the time $T_B$= 30 ns (i.e. the QD is not filled with certainty with an additional electron at the end of this pulse stage), which leads to an over-estimate for the rate γ.

Figure S3 shows similar measurements performed on different resonances. Also in this case there is a good agreement between the energies of the excited states extracted from finite-bias DC spectroscopy and pulse-gate spectroscopy measurements.

### 3. Transient current through excited states

In Figure S4 we show measurements analogous to the one presented in Fig. 4 in the main text, and performed in the very same sample discussed there, but in different regimes for the tunnel coupling $\Gamma_L$, $\Gamma_R$. In all cases saturation of the number of electrons transmitted via an excited state is observed, with a characteristic rate $\gamma$ that, as expected, depends on the barrier-gate configuration.

s1. Ferrari, A. C. *et al.* Raman Spectrum of Graphene and Graphene Layers. *Phys. Rev. Lett.* **97**, 187401 (2006).

s2. Stampfer, C., Schurtenberger, E., Molitor, F., Güttinger, J., Ihn, T., Ensslin, K. Tunable graphene single electron transistor. *Nano Lett.* **8**, 2378 (2008).

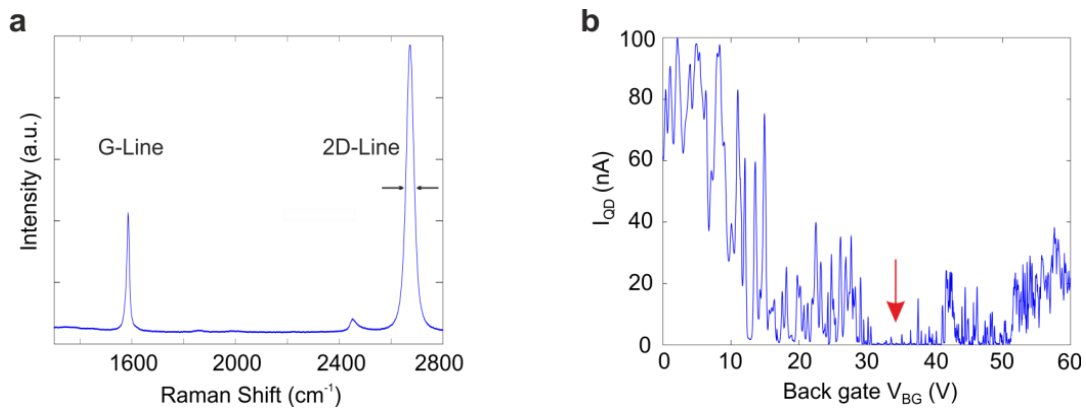

Figure S1: **Sample characterisation. a.** Raman spectrum of the measured graphene flake, recorded directly after exfoliation. **b.** Back gate characteristics of our quantum dot, measured at an electron temperature below 100 mK and bias voltage $V_{SD} = 15$ mV. For $V_{BG} < 31$ V and $V_{BG} > 41$ V, the device is in the hole- and in the electron-transport regime, respectively. In both cases, a significant current can flow through the dot. Vice versa, for $31$ V $< V_{BG} < 41$ V the device is in the transport gap and the current through the dot is strongly suppressed. The arrow indicates the back gate voltage $V_{BG} = 34.6$ V at which all other measurements have been performed.

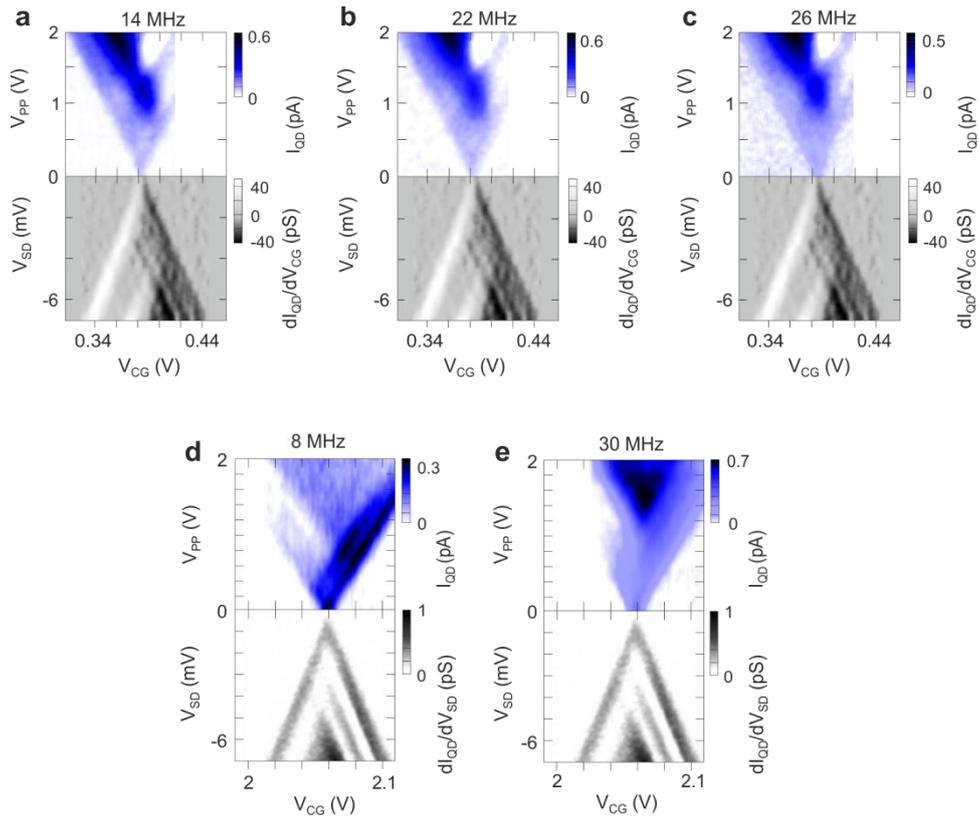

Figure S2: **Pulsed-gate spectroscopy of excited states at different frequencies.** These data complement those presented in Fig. 3 in the main text. Panels **a-c** refer to the resonance shown in Figs. 3g,h, measured at frequencies of 14, 22 and 30 MHz, respectively. Panels **d,e** refer to the resonance shown in Figs. 3b-e, measured at frequencies of 8 and 30 MHz. In all cases, upper and lower half of the panels correspond to pulsed-gate and to finite-bias spectroscopy, respectively.

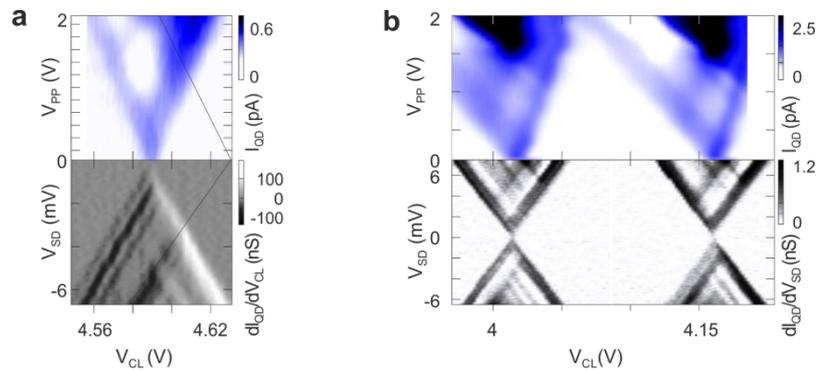

Figure S3: **Pulsed-gate spectroscopy. a,b** Pulsed-gate spectroscopy of different resonances (pulse frequency 25 MHz, $V_{SD}$ = -1 mV) and comparison of the results with corresponding DC measurements are shown. In panel **b** two neighbouring states are shown.

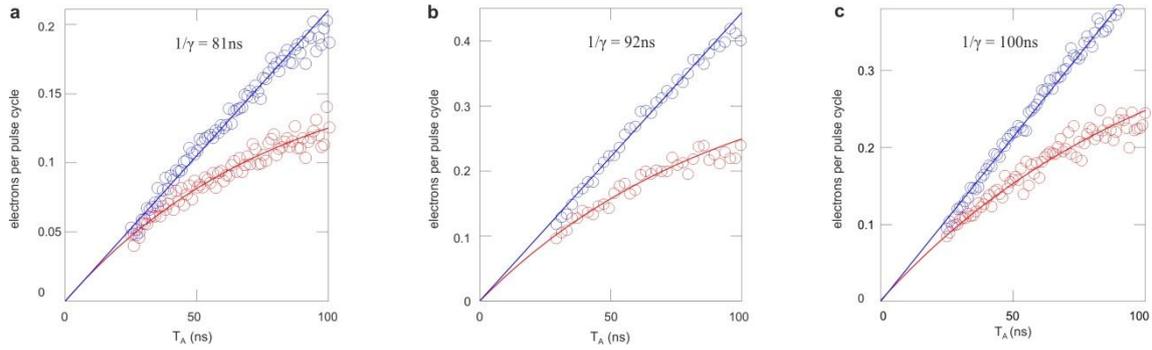

Figure S4: **Transient current through excited states.** Measurements of the transient dynamics of transport via excited states along the scheme of Fig. 4 in the main text. The three plots correspond to different voltages applied to the barrier-gate. In all measurements, the number of electrons transferred via the ground state increases linearly with the duration of the lower pulse level $T_A$, while the number of electrons transmitted via the excited state tends to saturate. The characteristic time constant $1/\gamma$ of this saturation process as well as the asymptotic number of transmitted electrons $n_{sat}$ varies when changing the voltages at the barrier-gates, as these affects the tunnelling rates. **a.** Voltage applied to the barrier-gates: $V_{RG} = 50$ mV and $V_{LG} = 420$ mV. Fit parameters extracted from the data relative to transport via excited states: $n_{sat} = 0.176$ and $1/\gamma = 81$ ns. **b.** Barrier-gates voltage: $V_{RG} = 0$ mV and $V_{LG} = 410$ mV. Fit parameters: $n_{sat} = 0.376$ and $1/\gamma = 92$ ns. **c.** Barrier-gates voltage: $V_{RG} = 0$ mV and $V_{LG} = 400$ mV. Fit parameters: $n_{sat} = 0.314$ and $1/\gamma = 100$ ns.